\documentclass[reprint,cite,doublecol,figures,prb,amsmath,amssymb,superscriptaddress,showpacs]{revtex4-1}
\usepackage[dvips]{graphicx}
\DeclareGraphicsExtensions{.pdf} \graphicspath{{./figs/}}
\usepackage{color}
\usepackage{verbatim}
\usepackage[german,english]{babel}

\usepackage{dcolumn}% Align table columns on decimal point
\usepackage{bm}% bold math

\newcommand{\CuDCl}{2(1,4-Dioxane)$\cdot$2(H$_2$O)$\cdot$CuCl$_2$}

\newcommand{\NSM}{Neutron Scattering and Magnetism, Laboratory for Solid State Physics, ETH Z\"urich, CH-8093 Z\"urich, Switzerland}

\newcommand{\figref}[1]{Fig.~\,\protect\ref{#1}}

\begin{document}

%%%\preprint{APS/123-QED}

\title{Finite-temperature scaling of spin correlations in a partially magnetized Heisenberg $S=1/2$ chain}
\author{M.~H\"alg}
\thanks{Corresponding author}
\email{haelgma@phys.ethz.ch}
\affiliation{\NSM}
\author{D.~H\"uvonen}
\altaffiliation{Previous address: Neutron Scattering and Magnetism, Laboratory for Solid State Physics, ETH Z\"urich, CH-8093 Z\"urich, Switzerland.}
\affiliation{National Institute of Chemical Physics and Biophysics, 12618 Tallinn, Estonia}
\author{N.~P.~Butch}
\affiliation{Center for Neutron Research, National Institute of Standards and Technology, MS 6100 Gaithersburg, Maryland 20899, USA}
\author{F.~Demmel}
\affiliation{ISIS Facility, Rutherford Appleton Laboratory, Chilton, Didcot, Oxon OX11 0QX, United Kingdom}
\author{A.~Zheludev}
\homepage{http://www.neutron.ethz.ch/}
\affiliation{\NSM}

\date{\today}

\begin{abstract}
Inelastic neutron scattering is employed to study transverse spin correlations of a Heisenberg $S=1/2$ chain compound in a magnetic field of 7.5~T.
The target compound is the antiferromagnetic Heisenberg $S=1/2$ chain material \CuDCl, or CuDCl for short.
The validity and the limitations of the scaling relation for the transverse dynamic structure factor are tested, discussed and compared to the Tomonaga-Luttinger spin liquid theory and to Bethe-ansatz results for the Heisenberg model.

\end{abstract}

\maketitle

\section{\label{sec:Introduction}Introduction}

For interacting fermions in one dimension, the notion of Landau's Fermi liquid breaks down.
A new theoretical framework, known as the Tomonaga-Luttinger liquid (TLL), takes it's place.\cite{Tomonaga1950,Luttinger1963,Tsvelikbook,Giamarchibook}
The TLL is a quantum critical state, since quantum fluctuations dominate down to lowest temperatures, $T \rightarrow 0$.\cite{Sachdevbook}
Hence, its properties follow universal laws.\cite{Sachdevbook}
The effective low-energy Hamiltonian is written in a universal form with only two key parameters, namely the dimensionless Luttinger parameter $K$, which quantifies the interaction strength between fermions,  and the Fermi velocity $u$, which links energy and momentum of the low-energy dynamics.
The Luttinger parameter value $K=1$ describes free fermions, while $K<1$ and $K>1$ correspond to repulsive and attractive interactions, respectively.\cite{Giamarchibook}
The TLL theory fully describes the low-energy properties, including all thermodynamics and correlation functions, for a variety of entirely different microscopic one-dimensional (1D) models.\cite{Haldane1980,Haldane1981,Giamarchibook}
Experimentally, the TLL has been realized in quasi-1D conductor materials\cite{Denlinger1999,Wang2008}, fractional quantum Hall fluids\cite{Chang1996,Grayson1998}, or quantum wires, e.g. carbon nanotubes\cite{Ishii2003,Rauf2004}.

Some of the best realizations of the TLL are found in certain 1D quantum spin systems, including spin chains and ladders.\cite{Tsvelikbook,Giamarchibook}
Since spins are the relevant degrees of freedom in this case, they are referred to as Tomonaga-Luttinger spin liquids (TLSL).
The humble  antiferromagnetic (AF) Heisenberg $S=1/2$ chain is one of the oldest and best known examples.
Real materials described by this model have provided a convenient platform for the study of TLSL physics using e.g. thermodynamic measurements,\cite{Yoshida2005,Umegaki2011,Umegaki2012,Haelg2014,Kono2015} neutron spectroscopy,\cite{Denderthesis,Lake2005,Zheludev2007} or nuclear magnetic resonance (NMR) experiments \cite{Sirker2009,Kuehne2011}.
One attraction of spin systems is that their fermionic interactions are continuously tunable by applying an external magnetic field.
This feature has been successfully exploited in several recent experimental studies.\cite{Kuehne2011,Jeong2013,Povarov2015}
For the simple AF Heisenberg $S=1/2$ chain, the Luttinger parameter is expected to vary continuously from $K=0.5$ in zero field to $K=1$ at saturation.\cite{Giamarchibook}
This implies that the low-energy spin correlation functions, which are universally expressed in terms of $K$ and $u$, become field-dependent as well.
Neutron scattering experiments directly probe these correlation functions, and can thus be used to study the evolution of the TLSL in applied fields.

To date, neutron scattering measurements of TLSL correlation functions in AF Heisenberg $S=1/2$ chains have only been performed in the absence of a magnetic field.\cite{Denderthesis,Lake2005,Zheludev2007}
The main result is that the AF dynamic structure factor $S(\pi,\omega)$ and the local dynamic structure factor $S(\omega)$ can be written as universal scaling forms of $\omega/T$, with the scaling functions explicitly given by TLSL theory with $K=0.5$.
As discussed below, extending such measurements to the case of applied magnetic fields and consequently larger $K$ values, presents unique experimental challenges. In the present work we attempt to overcome these difficulties.
We perform finite-temperature measurements of $S(\pi,\omega)$ in the spin chain compound \CuDCl{}, or CuDCl for short, in a magnetic field of roughly half the saturation value.
Despite a limited dynamic range, we find that the measured dynamic structure factor does indeed show finite-temperature scaling in agreement with TLSL theory.

\section{Experimental considerations \label{sec:challenges}}

\begin{figure}[!htb]
\unitlength1cm
\includegraphics[width=.48\textwidth]{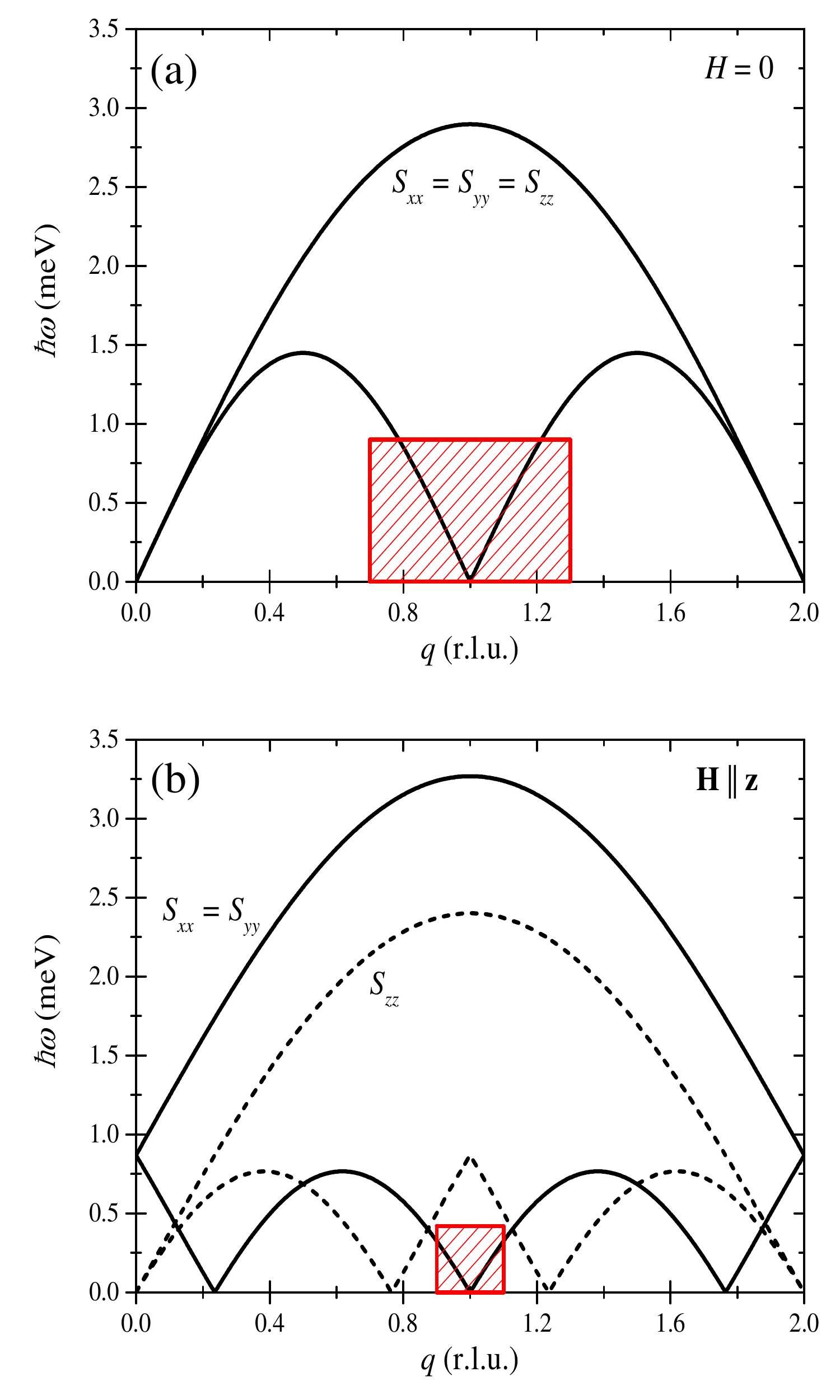}
\caption{A sketch of the spin excitation spectrum of an AF Heisenberg $S=1/2$ chain in zero magnetic field (a) and with magnetic field applied along the $z$ axis, as in the experiment (b). The numerical values correspond to the values of the target material of the present work, CuDCl. The approximate region where correlations are governed by TLSL physics is hatched. \label{fig:theory}}
\end{figure}

In  the absence of an external magnetic field, probing TLSL physics in $S=1/2$ Heisenberg AF materials using inelastic neutron scattering is comparatively straightforward.
At low temperatures, the excitation spectrum is a multi-spinon continuum with a sharply defined lower bound.\cite{Mueller1981}
However, only a part of these correlations are actually described by the TLSL theory,\cite{Lake2005} since it relies on a linear dispersion of excitations.
Figure~\ref{fig:theory}(a) shows the spinon continuum and the approximate region in energy-momentum space, where the TLSL predictions for the dynamic structure factor are applicable.
This region is relatively large, leaving a wide dynamic range available for investigations.
The spin structure factor is generally polarization-dependent, being defined as:
\begin{equation}
S_{\alpha\alpha}\left(\textbf{q}, \omega\right) = \frac{1}{2\pi\hbar}\int{\left\langle S^{\alpha}\left(\textbf{x},t\right)S^{\alpha}\left(0,0\right) \right\rangle e^{-i\left(\textbf{qx}-\omega t\right)}dtd\textbf{x}}.
\end{equation}
In zero applied field, spin fluctuations are isotropic and, thus, $S_{\alpha\alpha}\left(\textbf{q}, \omega\right)$ is the same for all spin components $\alpha$.
Therefore, there is no need to discriminate between different polarization channels in a neutron-scattering experiment, which greatly facilitates the measurements.

In a nonzero external magnetic field, the constraints on probing TLSL physics with neutron scattering are much more severe.
A field along the $z$ axis breaks the full rotational symmetry, so that the spin correlations become anisotropic, $S_{xx}\left(\textbf{q}, \omega\right)=S_{yy}\left(\textbf{q}, \omega\right)\ne S_{zz}\left(\textbf{q}, \omega\right)$.
The total spectrum now consists of distinct and overlapping longitudinal and transverse contributions.
As shown in  Figure~\ref{fig:theory}(b), each of these is a continuum.
The corresponding sharp lower bounds are now distinguishable, with incommensurately positioned minima.\cite{Mueller1981,Giamarchibook}
The incommensurability is directly proportional to the field-induced magnetization, so that the minima of the spectrum are shifted from the zero-field commensurate positions by $\delta q \sim 2\pi\left\langle S^z \right\rangle = \frac{2\pi}{N}\sum_{i=1}^N\left\langle S^z_i \right\rangle$ at $T\rightarrow 0$.\cite{Mueller1981,Giamarchibook}
Such a spectacular restructuring of the excitation spectrum has been confirmed experimentally by neutron-scattering studies, including those on the well-known spin-chain material CuPzN.\cite{Stone2003}
In application to the present problem, the consequences of this spectral changes are three-fold.
First, the lower bound of either continuum is lowered compared to the one in zero field and, thus, the linear-dispersion region where the TLSL notion may apply is reduced.
Secondly, the TLSL theory predicts different scaling forms for the longitudinal and transverse components near their respective minima.
While it is in principle possible to discriminate between different polarization channels using polarized neutrons, in practice the corresponding intensity penalty makes the experiment exceedingly complicated or even infeasible.
The only solution is to further shrink the measurement window to totally avoid the overlap between continua of different polarizations, as illustrated in Figure~\ref{fig:theory}(b).
Thirdly, only considering one polarization channel at least halves the net intensity compared to the zero-field case where the scattering of all polarization channels can be exploited.

One way to overcome these obstacles is to increase the energy range by using a target spin-chain compound with a larger exchange constant $J$.
However, substantially magnetizing such a material would require  unattainably large magnetic fields.
The alternative is to use a compound with a small $J$, but to collect the data with tight energy and momentum resolutions. Unfortunately, resolution always comes at the expense of intensity.
The answer to this dilemma is to make use of very large single-crystal samples, ideally as large as the neutron beam itself. This is the approach adopted in the present study.

\begin{figure}[!htb]
\unitlength1cm
\includegraphics[width=.48\textwidth]{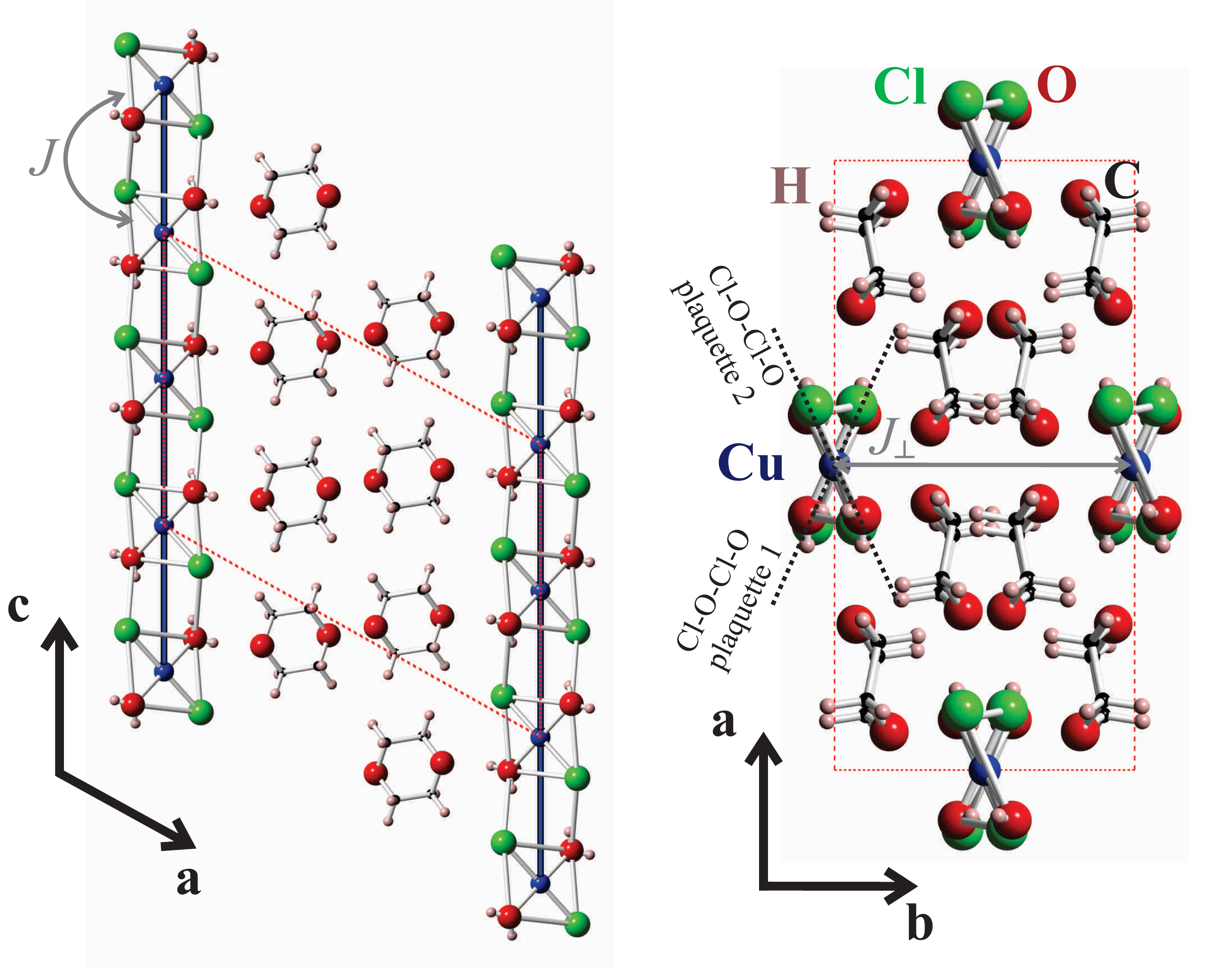}
\caption{Crystal structure of CuDCl. The magnetic Cu$^{2+}$ ions are linked by Cl$^-$ ions and water molecules and form chains along $c$ with coupling constant $J$. Individual chains are separated by 1,4-Dioxane molecules. The crystallographic unit cell is indicated by dotted lines. The Cu$^{2+}$ ions are surrounded by Cl-O-Cl-O plaquettes that are slightly tilted relative to one another in a staggered arrangment. The strongest interchain coupling, $J_{\perp}$, is expected to be along the $b$ axis.
\label{fig:structure}}
\end{figure}

Guided by the potential of growing very large single crystals, for our experiments we selected the Heisenberg $S=1/2$  chain material CuDCl.\cite{Hong2009}
It has a monoclinic structure, space group $C2/c$, and lattice constants $a = 17.43$\AA, $b=7.48$~\AA, $c=11.82$~\AA, and $\beta = 119.4^{\circ}$ at 173~K.\cite{Hong2009}
The spin chains run along the crystallographic $c$ axis and the leading term in the Hamiltonian is the nearest-neighbor Heisenberg exchange with exchange constant $J\sim 0.9~\text{meV}$.\cite{Hong2009}
The chains are formed by Cu$^{2+}$ ions, which are surrounded by two O$^{2-}$ and Cl$^{-}$ ions (Fig.~\ref{fig:structure}).
The resulting Cl-O-Cl-O plaquettes are slightly tilted with respect to the neighboring ones in the same chain, resulting in a staggered arrangement.
The individual chains are well separated from each other by 1,4-Dioxane molecules along the $a$ and $b$ axes, so that the interchain interactions are expected to be very small.
To date, no magnetic ordering has been reported in this system.

\section{Experimental details}
All experiments were performed on fully-deuterated single-crystal samples of CuDCl.
The crystals were grown in a nitrogen atmosphere from a methanol solution containing water, anhydrous CuCl$_2$ and 1,4-Dioxane.
The crystals are highly unstable in ambient air.
The 1,4-Dioxane molecules evaporate and leave CuCl$_2$ powder if the CuDCl crystal is not stabilized by an overpressure of 1,4-Dioxane in the atmosphere or by an applied pressure.

Specific-heat data were collected on a commercial Quantum Design physical property measurement system (PPMS) using a Quantum Design dilution insert for measurements at lowest temperatures down to 50~mK.
The sample was covered with Apiezon N grease in order to prevent the 1,4-Dioxane molecules from evaporating.
Before the sample was mounted, the specific heat of the used Apiezon N was measured in order to subtract its contribution from the total specific heat.
Each data point was measured with a temperature rise of 2~$\%$ of the sample temperature.

Inelastic-neutron-scattering experiments were performed on the time-of-flight spectrometers OSIRIS at ISIS and DCS at the NIST Center for Neutron Research (NCNR).
Fully-deuterated single crystals of mass 11~g (on OSIRIS) and 12.5~g (on DCS), respectively, were used.
The samples were sealed in He atmosphere and a small amount of deuterated 1,4-Dioxane was added to generate an overpressure of 1,4-Dioxane inside the sample containers.
The crystals were aligned with the incident neutron beam in the $ac$ plane.
A 7.5~T vertical magnet ($\textbf{H}~||~\textbf{b}$) was employed on OSIRIS and a 10~T vertical magnet was used on DCS.
Both experiments were performed with dilution inserts for the magnets.
The final neutron wavelength was set to 6.66~$\text{\AA}$ on OSIRIS.
On DCS an incident neutron wavelength of 8.00~$\text{\AA}$ was used.
The experimental energy resolution, defined as the full width at half maximum (FWHM) of the elastic incoherent scattering was 0.03~meV on OSIRIS and 0.06~meV on DCS.

\section{Experimental results}
\begin{figure}[!htb]
\unitlength1cm
\includegraphics[width=.48\textwidth]{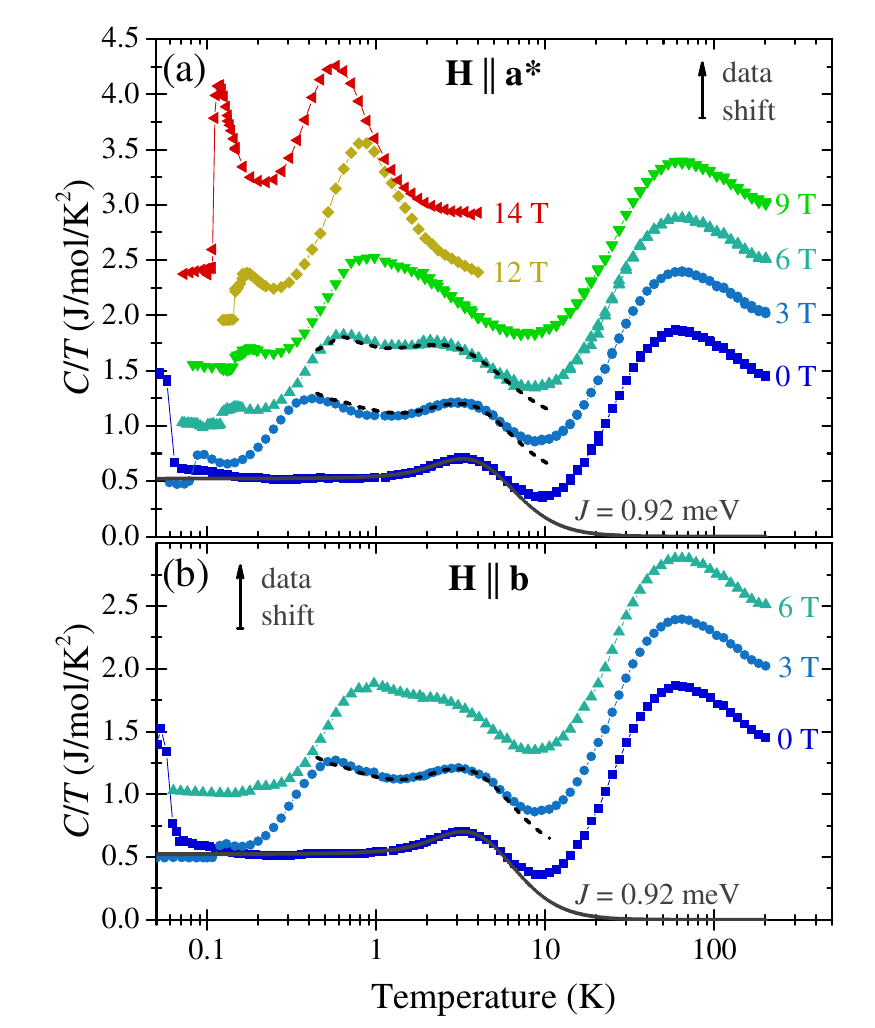}
\caption{Symbols: measured specific-heat of CuDCl for various magnetic fields applied perpendicular to the chain direction. An offset of 0.5~J/mol/K$^2$, illustrated by an arrow in the graph, is added between the data sets for visibility. The solid line is the theoretical result for an AF Heisenberg $S=1/2$ chain in zero field with $J=$~0.92~meV, as explained in the text. Dashed lines are DMRG data taken from reference~\onlinecite{Shibata2001} for a Heisenberg $S=1/2$ chain with a staggered magnetic field of approximately 0.1~T. \label{fig:cmol}}
\end{figure}

Specific heat of CuDCl was measured with a magnetic field applied perpendicular to the chain direction.
Respective data for a magnetic field along the $a^{\ast}$ and $b$ axes are displayed in \figref{fig:cmol}.
In zero magnetic field three features can be clearly discerned.
At high temperatures ($T>$~10~K) the contribution from phonons is dominant.
The data below 10~K, in particular the maximum at about 3~K, are in very good agreement with the theoretical results for an AF Heisenberg $S=1/2$ chain from Johnston \textit{et al.}\cite{Johnston2000} with $J=0.92~\text{meV}$ as the only free parameter (solid line in \figref{fig:cmol}).\footnote{Eqs. (54a-c) from reference~\onlinecite{Johnston2000}:\\ $C_{\text{mol}} = \text{N}_{\text{A}}\text{k}_{\text{B}}\left(\frac{3}{16t^2}\mathcal{P}_{(9)}^{(6)}\left(t\right)-F\left(t\right)\right)$,\\
 $\mathcal{P}_{(9)}^{(6)}\left(t\right) = \frac{1+\sum_{n=1}^6 N_n/t^n}{1+\sum_{n=1}^9 D_n/t^n}$,\\
 $F\left(t\right) = a_1t\sin\left(\frac{2\pi}{a_2+a_3t}\right)e^{-a_4t}+a_5te^{-a_6t}$,\\
 where $t=\text{k}_{\text{B}}T/J$, $\text{N}_{\text{A}}$ is the Avogadro constant and $N_n$, $D_n$ and $a_n$ are coefficients given in the reference.}
The constant low-temperature part of the $C/T$  curve spans over more than one decade in temperature and is characteristic of the TLSL phase.
Below 100~mK, at the lowest accessible temperatures, the specific heat increases abruptly, suggesting the onset of three-dimensional magnetic long-range order.

In applied fields, the ordering peak moves to higher temperatures, which is consistent with an expected increase of the ordering temperature $T_\mathrm{N}$ in weakly coupled spin chains.\cite{DeGroot1986,Wessel2000,Zvyagin2010}
However, a new maximum emerges around 0.3~K which is \textit{not} a feature of ideal 1D Heisenberg $S=1/2$ antiferromagnets.\cite{Bonner1964}
At higher fields, this new maximum merges with the one initially seen around 3~K.

\begin{figure}[!htb]
\unitlength1cm
\includegraphics[width=.48\textwidth]{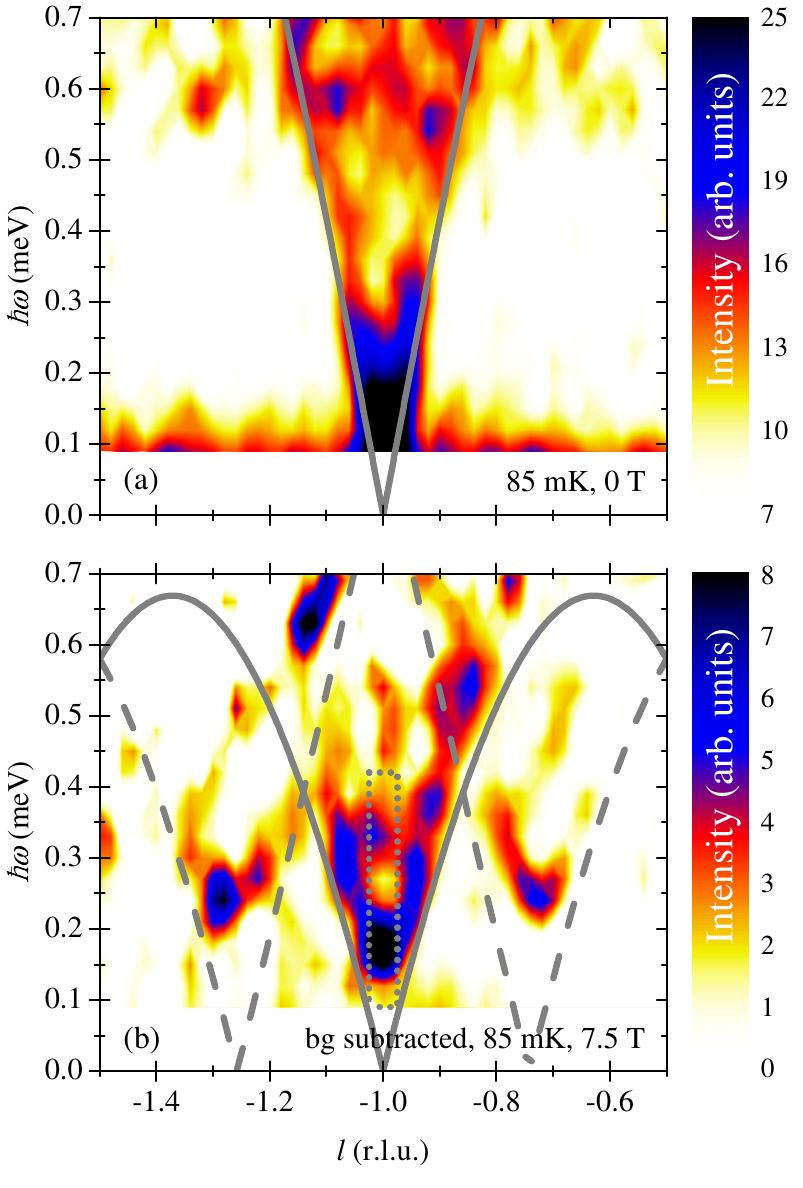}
\caption{False-color representation of the inelastic neutron spectra  measured in CuDCl on DCS at 85~mK in zero magnetic field (a) and at 7.5~T (b). The solid and dashed lines represent the lower boundaries of the continua for transverse (solid) and longitudinal (dashed) excitations in an ideal AF Heisenberg $S=1/2$ chain compound with $J=0.92$~meV at 7.5~T. The dotted rectangle indicates the region of interest for the present study. The left and the right boundary of the rectangle mark the limits of the integration for obtaining $S\left(\pi,\omega\right)$. \label{fig:spectrum}}
\end{figure}

The neutron spectra of CuDCl were first measured on both spectrometers in zero magnetic field at base temperature.
Figure~\ref{fig:spectrum}(a) is a false-color plot of the neutron intensities collected on DCS in the vicinity of the 1D AF zone-center  at 85~mK.
These data were taken at several sample orientations, then projected along the $a^{\ast}$ axis onto the relevant $(l,\hbar \omega)$ plane.
The result agrees well with previous neutron studies.\cite{Hong2009}

\begin{figure}[!htb]
\unitlength1cm
\includegraphics[width=.48\textwidth]{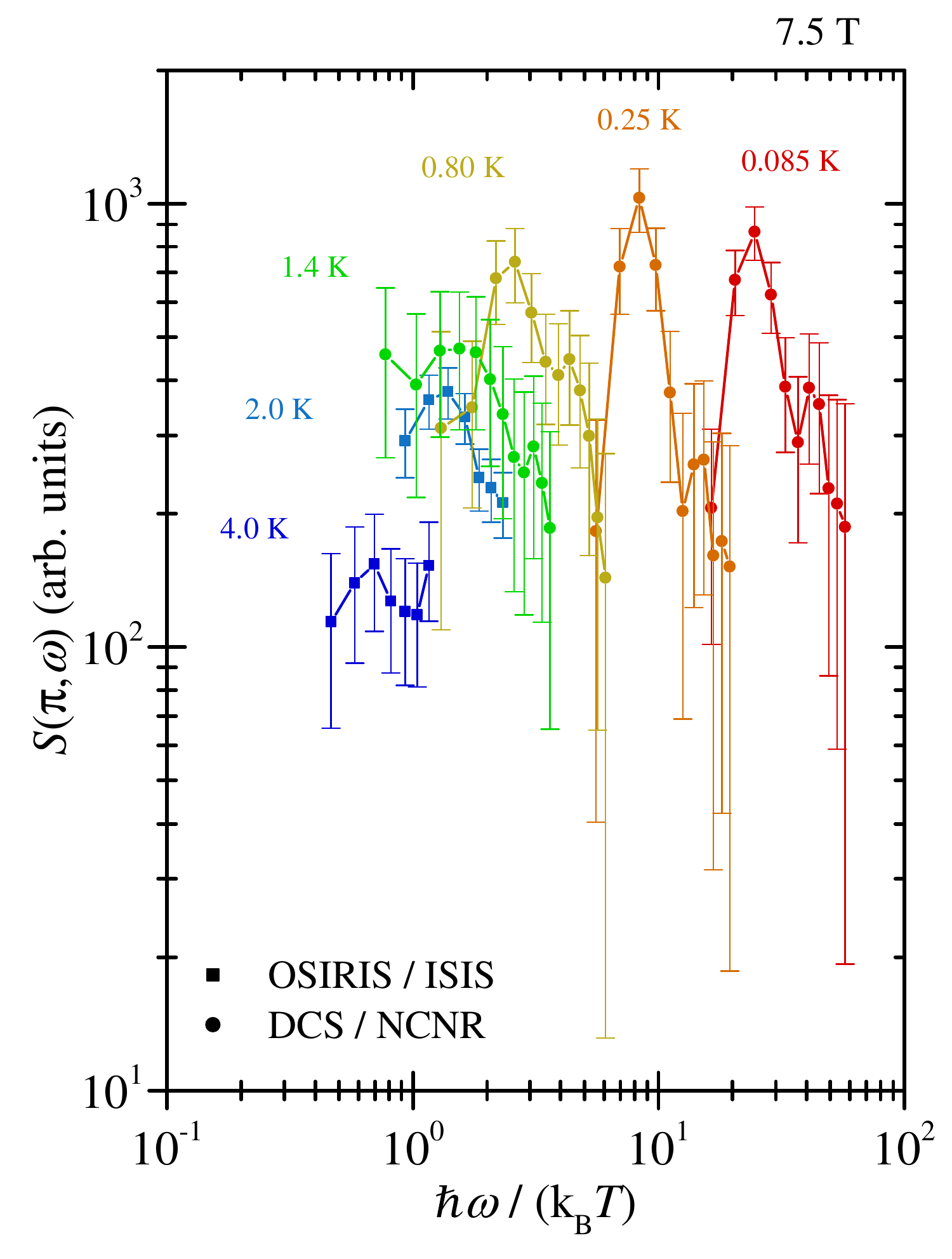}
\caption{Dynamic structure factor at the AF wave vector, $l=-1$, versus the temperature-scaled energy $\hbar\omega/(\text{k}_{\text{B}}T)$ for all neutron-scattering data obtained on OSIRIS and DCS at 7.5~T for $\hbar\omega \leq$~0.42~meV. The energy gap is clearly visible below 1~K as a drop of intensity in the low-energy tail of $S\left(\pi,\omega\right)$. The error bars correspond to the statistical errors of the neutron-scattering measurements. \label{fig:unscaled}}
\end{figure}

The spectra in a magnetic field of 7.5~T were collected at 2.0~K and 4.0~K on OSIRIS, as well as at 0.085~K, 0.25~K, 0.80~K and 1.4~K on DCS.
The data taken at low temperatures in zero field were used to determine the background for the high-field experiments.
For each energy transfer, a constant background was fitted to the part of the zero-field data which is not affected by the continuum of the magnon excitations, i.e. which lies outside of the boundaries indicated by the solid lines in \figref{fig:spectrum}(a).
The resulting background-subtracted spectrum for 7.5~T and 85~mK obtained on DCS is depicted in \figref{fig:spectrum}(b).
As expected for Heisenberg $S=1/2$ chains, the spin correlations parallel and transverse to the applied magnetic field show minima at different wave vectors.
Based on the known values for the coupling constant $J$ and the magnetic field $H$, and assuming a $g$ factor $g=2$, the expected boundaries of the continua for transverse and longitudinal excitations are shown as solid and dashed lines, respectively.
The longitudinal excitations have an energy minima at $l=-1\pm0.27$, which corresponds to $q=\pi\pm 0.85$ for the structure of CuDCl.
From this figure it can be seen that at the 1D AF zone center, $l=-1$, at low energies, one only measures the transverse spin correlations, corresponding to $S_{xx}(\pi,\omega)$.
Integrating the experimental data in a narrow range $-1.025\le l \le -0.975$ [dotted rectangle in \figref{fig:spectrum}(b)] yields the measured energy dependence of this quantity, summarized in \figref{fig:unscaled} for different temperatures and both experiments.

From \figref{fig:spectrum}(b) it is immediately obvious that an applied magnetic field opens an energy gap in the spin excitation spectrum, even though at zero field it is gapless.
The energy gap is approximately $\Delta\sim 0.18~\text{meV}$ at 7.5~T.
The behavior is very similar to the one previously studied in the spin-chain compound CDC.\cite{Kenzelmann2004,Kenzelmann2005}
In that case, it was attributed to a staggered spin field due to a combination of a uniform magnetic field and a staggered gyromagnetic tensor in the spin chains.\cite{Oshikawa1997,Kenzelmann2004,Kenzelmann2005,Umegaki2012}
For CuDCl, a staggering of the $g$ tensor would be a natural consequence of the above-mentioned staggered tilting of the Cl-O-Cl-O plaquettes.
A staggered $g$ tensor also accounts for the observed double-maximum shape of the $C/T$ curves in CuDCl.
The measured data are in good qualitative agreement with previous density-matrix renormalization-group (DMRG) calculations for the Heisenberg $S=1/2$ chain in a staggered magnetic field (dashed lines in \figref{fig:cmol}).\cite{Shibata2001}

\section{Discussion\label{sec:discussion}}

The ideal Heisenberg $S=1/2$ chain below saturation, being in the TLSL state, is always quantum critical.\cite{Sachdevbook}
However, the opening of the gap $\Delta$ in the excitation spectrum is a serious complication in the quest to study quantum critical properties in CuDCl.
Indeed, the symmetry-breaking staggered field produced in CuDCl by a uniform applied field takes the system away from the TLSL criticality.
But even in this case, TLSL physics may be accessible in the quantum critical regime at elevated temperatures or on short time scales.\cite{Sachdevbook}
Specifically, quantum critical fluctuations dominate for
\begin{equation}
\left(\text{k}_{\text{B}}T \gg \Delta\right) \text{ or } \left(\hbar\omega \gg \Delta\right).
\end{equation}
With this in mind, in all the subsequent analysis we only used those data points for which either the temperature or the energy transfer are much larger than the gap $\Delta$, i.e.
\begin{equation}
\left(T \geq 2.8~\text{K}\right) \text{ or } \left(\hbar\omega \geq 0.24~\text{meV}\right),
\label{eq:condition1}
\end{equation}
where the numerical limits are chosen as the measured energy resolution on DCS added to the gap energy.
An additional restriction is set by the required linearity of the dispersion, as discussed in section~\ref{sec:challenges}.
In the present case for 7.5~T [see plot of the lower continuum boundary in \figref{fig:theory}(b) and \figref{fig:spectrum}(b)], the upper limit of the energy transfer and the temperature was chosen to be:
\begin{equation}
\text{k}_{\text{B}}T,\hbar\omega \lesssim 0.4~\text{meV}.
\label{eq:condition2}
\end{equation}

Finally, another effect which can potentially drive the system away from TLSL quantum criticality is three-dimensional interchain coupling.
Fortunately, it is negligibly small in CuDCl for the temperatures and energy transfers defined above ($\text{k}_{\text{B}}T_{\text{N}}|_{\text{7.5 T}} \sim 0.02~\text{meV} \ll ~\Delta$, cf. Fig.~\ref{fig:cmol}).
Indeed, from the zero-field ordering temperature determined from the specific-heat data, $T_{\text{N}} \lesssim 55~\text{mK}$, the interchain coupling can be estimated to be $J_{\perp} \lesssim 4$~$\mu$eV$\ll \Delta$, following reference~\onlinecite{Yasuda2005}.

\begin{figure}[!htb]
\unitlength1cm
\includegraphics[width=.48\textwidth]{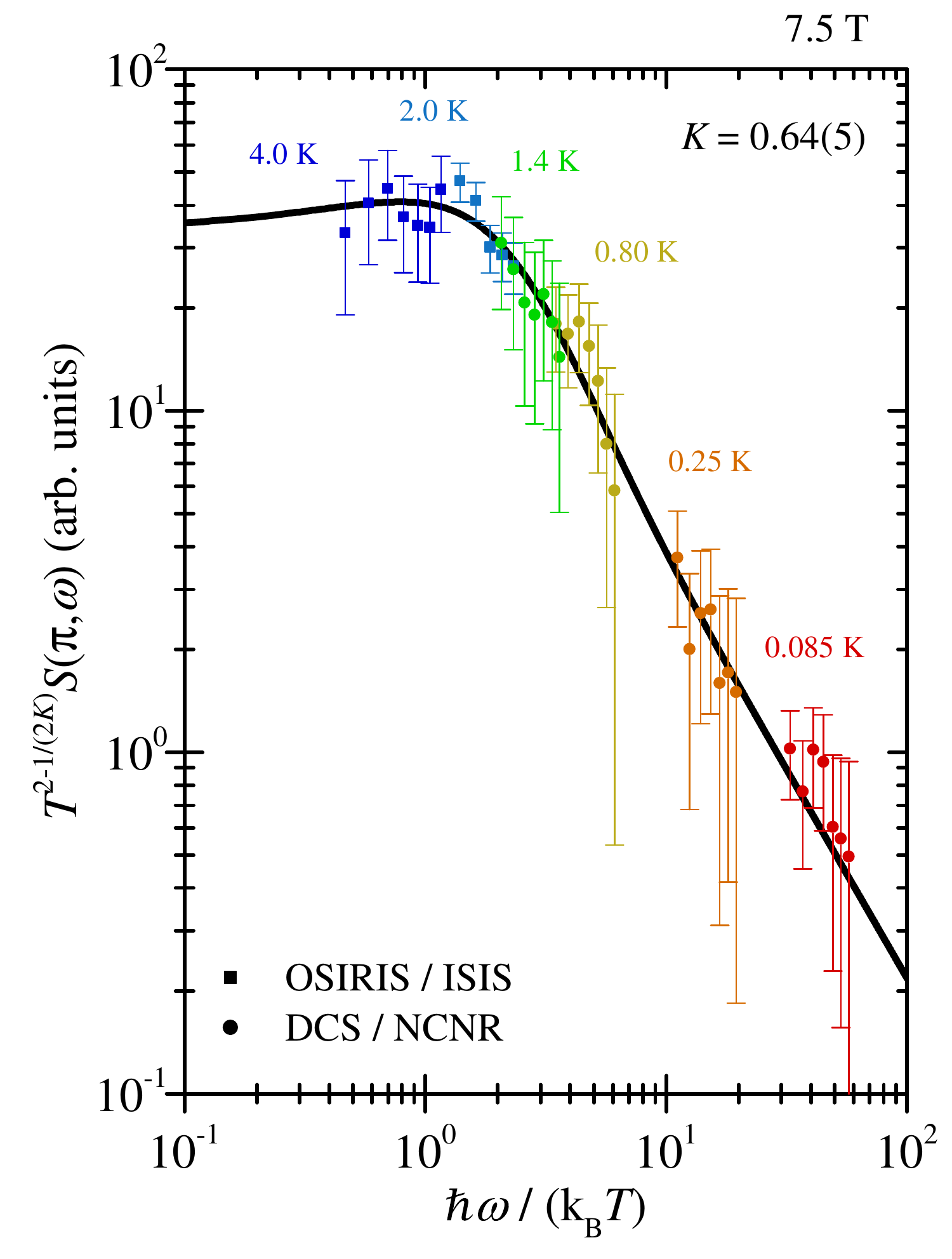}
\caption{Scaling plot for the Luttinger parameter $K=0.64$. The solid line shows the universal scaling function given in eq.~(\ref{eq:phi}). The error bars correspond to the statistical errors of the neutron-scattering measurements. \label{fig:scaled}}
\end{figure}

In general, the correlation function at the critical wave vector $q=\pi$ is expected to have the following scaling form in the quantum critical regime:
\begin{equation}
S\left(\pi,\omega \right) = T^{-\alpha}\Phi\left(\frac{\omega}{T}\right),
\label{eq:s}
\end{equation}
with the scaling exponent $\alpha$ and the scaling function $\Phi$.
Quantum criticality implies that the microscopic Hamiltonian parameters $J$ or $H$ do not enter this expression explicitly.
The only relevant energy scale is set by the temperature itself.
While the narrow dynamic range and the small number of different measurement temperatures prevent the confirmation of such a scaling law in a model-free fashion -- as was done in our recent studies on magnetized spin ladders,\cite{Povarov2015} or anisotropic spin chains at the Ising quantum critical point,\cite{Haelg2015a} where the experimental considerations were less restrictive -- the CuDCl data are sufficient to verify the specific prediction for the scaling function $\Phi$ and the scaling exponent $\alpha$ provided by the TLSL theory:\cite{Schulz1986,Giamarchibook}
\begin{eqnarray}
\Phi\left(\frac{\omega}{T}\right) & \propto	& \frac{1}{1-e^{-\frac{\hbar\omega}{\text{k}_{\text{B}}T}}} \nonumber \\
	&	 \times & \text{Im}\left[\left(\frac{\Gamma\left(\frac{1}{8K}-i\frac{\hbar\omega}{4\pi\text{k}_{\text{B}}T}\right)\Gamma\left(1-\frac{1}{4K}\right)}{\Gamma\left(1-\frac{1}{8K}-i\frac{\hbar\omega}{4\pi\text{k}_{\text{B}}T}\right)}\right)^2\right],\label{eq:phi}
\\
\alpha &=& 2-\frac{1}{2K},
\label{eq:alpha}
\end{eqnarray}
where $\Gamma$ is Euler's gamma function.

Using only those $S(\pi,\omega)$ data points that satisfy the above-mentioned conditions on temperature and energy transfer, Eqs.~(\ref{eq:s}), (\ref{eq:phi}) and (\ref{eq:alpha}) were fit to the data.
The two only adjustable parameters were an overall constant scale factor and the Luttinger parameter $K$.
The best fit is obtained using $K=0.64(5)$.
The result is shown as a solid line in the scaling plot in \figref{fig:scaled}.
It can be seen that the data collected at different temperatures indeed align on a single curve with this choice of the scaling exponent.
Therefore, the validity of the universal scaling behavior of transverse spin correlations is verified.
The measured value of $K$ is clearly larger than $K=0.5$ for the unmagnetized chain.
Furthermore, it is in excellent agreement with the Bethe-ansatz result for AF Heisenberg $S=1/2$ chains with $J=0.92~\text{meV}$ at 7.5~T, which predicts $K \approx 0.65$.\cite{Giamarchibook}

\section{Conclusions}

Despite complications intrinsic to the selected prototype material, we have demonstrated that high-resolution neutron spectroscopy provides a means of probing the field-evolution of critical TLSL correlations in a partially magnetized Heisenberg $S=1/2$ chain antiferromagnet.
Furthermore, it was shown that the Luttinger parameter $K$ is increased in a magnetic field compared to the value obtained in the absence of a field in accordance with Bethe-ansatz calculations.

\section{Acknowledgments}
This project was supported by division II of the Swiss National Science Foundation.
Parts of this work are based on experiments performed at ISIS at the STFC Rutherford Appleton Laboratory in Oxfordshire, UK, and at the NIST Center for Neutron Research (NCNR) in Gaithersburg, Maryland, USA, supported in part by the National Science Foundation under Agreement No. DMR-0944772.
Additionally, this work was partly supported by the Estonian Ministry of Education and Research under grant No. IUT23-03 and the Estonian Research Council grant No. PUT451.
Identification of commercial materials or equipment does not imply recommendation or endorsement by the National Institute of Standards and Technology, nor does it imply that the materials or equipment identified are necessarily the best available for the purpose.

%%%%%%%%%%%%%%%%%%%%%%%%%%%%%%%%%%%%%%%%%%%%%%%%%%%%%%%%%%%%%%%%%%%%%%%%%%%%%%%%%

\bibliography{bibliography}\label{Cbib}

\end{document}